\documentclass[showpacs,twocolumn,floatfix,aps,prl,a4paper]{revtex4}
\usepackage{amsmath,amssymb,amsfonts,bm}

\usepackage{graphicx}
\usepackage{color}
\usepackage{bbold}

\newcommand{\bit}{\begin{itemize}}
\newcommand{\eit}{\end{itemize}}

\renewcommand{\>}{\right\rangle}
\newcommand{\<}{\left\langle}
\newcommand{\ba}{\begin{align}}
\newcommand{\ea}{\end{align}}
\newcommand{\be}{\begin{equation}}
\newcommand{\ee}{\end{equation}}
\newcommand{\bi}{\begin{itemize}}
\newcommand{\ei}{\end{itemize}}

\newcommand{\n}{\nonumber \\ }
\begin{document}

\newcommand{\bra}[1]{\< #1 \right|}
\newcommand{\ket}[1]{\left| #1 \>}

\title{Frustration and Correlations in Stacked Triangular Lattice Ising Antiferromagnets}
\author{F. J. Burnell$^{1,2}$ and J. T. Chalker$^1$}
\affiliation{$^1$Theoretical Physics, Oxford University, 1 Keble Road, Oxford OX1 3NP, United Kingdom}
\affiliation{$^2$School of Physics and Astronomy, University of Minnesota, Minneapolis, MN 55455, 
USA
}
\date{\today}

\begin{abstract}
We study multilayer triangular lattice Ising antiferromagnets with interlayer interactions that are weak and frustrated in an {\it abc} stacking. By analysing a coupled height model description of these systems, we show that they exhibit a classical spin liquid regime at low temperature, in which both intralayer and interlayer correlations are strong but there is no long range order. Diffuse scattering in this regime is concentrated on a helix in reciprocal space, as observed for charge-ordering in the materials LuFe$_2$O$_4$ and YbFe$_2$O$_4$.
\end{abstract}

\pacs{64.60.De, 
75.10.Hk,	
71.45.Lr	
}

\maketitle

Some simple models of highly frustrated magnets develop strong correlations at low temperature without long-range order. In this regime they are known as cooperative paramagnets or classical spin liquids \cite{balents}. The Ising model with nearest neighbour antiferromagnetic interactions on the triangular lattice was one of the earliest examples to be studied in detail \cite{tlafm}, while the Ising antiferromagnet on the pyrochlore lattice is an example of high current interest as a model for spin-ice materials \cite{spinicereview}. The description of 
the cooperative paramagnetic state 
presents a theoretical challenge, and may involve emergent degrees of freedom and fractionalized excitations: a height \cite{heightmodel} or gauge field \cite{gauge} in these cases, with vortex or monopole excitations.

In both these models the cooperative paramagnetic regimes
are continuously connected to disordered ground states with macroscopic entropy. By contrast, systems that have ordered ground states typically do not display strong correlations without long-range order, except near a critical point.
Here we study a remarkable exception: the three-dimensional Ising antiferromagnet built from {\it abc}-stacked triangular lattices, with interlayer coupling $J_{\perp}$ much weaker than the in-plane coupling $J$.
For $J_{\perp}=J$ this is the face-centered cubic lattice model, which orders discontinuously \cite{fcc} at a temperature $T_{\rm c} \simeq 1.74 J$, while for $J_\perp=0$ it reduces to uncoupled layers, which remain disordered to $T=0$. 
For $J_\perp \ll J$ we show that there is a temperature window in which the model has strong correlations, both between and within layers, and a large but finite correlation length. We formulate a theory for this regime in terms of coupled height fields. It is striking for its correlations (helical in reciprocal space), fluctuations (quartic in wavevector) and mechanism for suppression of long-range order (bound vortex-antivortex pairs). It also has interesting parallels with theories for smectic liquid crystals \cite{CL} and for frustrated quantum magnets \cite{Starykh}.

To put this layered antiferromagnet in context, note that the ordering pattern in magnets is determined at mean-field level by the location of minima in the eigenvalues of the exchange interaction matrix ${\rm J}({\bf q})$ as a function of wave vector $\bf q$. The suppression of ordering in many frustrated models is a consequence highly degenerate minima. As examples, nearest neigbour (nn) interactions on the pyrochlore lattice lead to a degenerate minimum band spanning the full, three-dimensional Brillouin zone \cite{gauge}, while competing first and second neighbour interactions on the diamond lattice give rise to minima on surfaces \cite{bergman}. On the {\it abc}-stacked triangular lattice, nn interactions  have long been noted for generating minima on {\it helical lines} in reciprocal space \cite{rastelli}. 
For the Ising model on this lattice we show in the following that reciprocal space correlations are concentrated close to such a helix over an extended temperature range.


 \begin{figure}
\includegraphics[width=.4\linewidth]{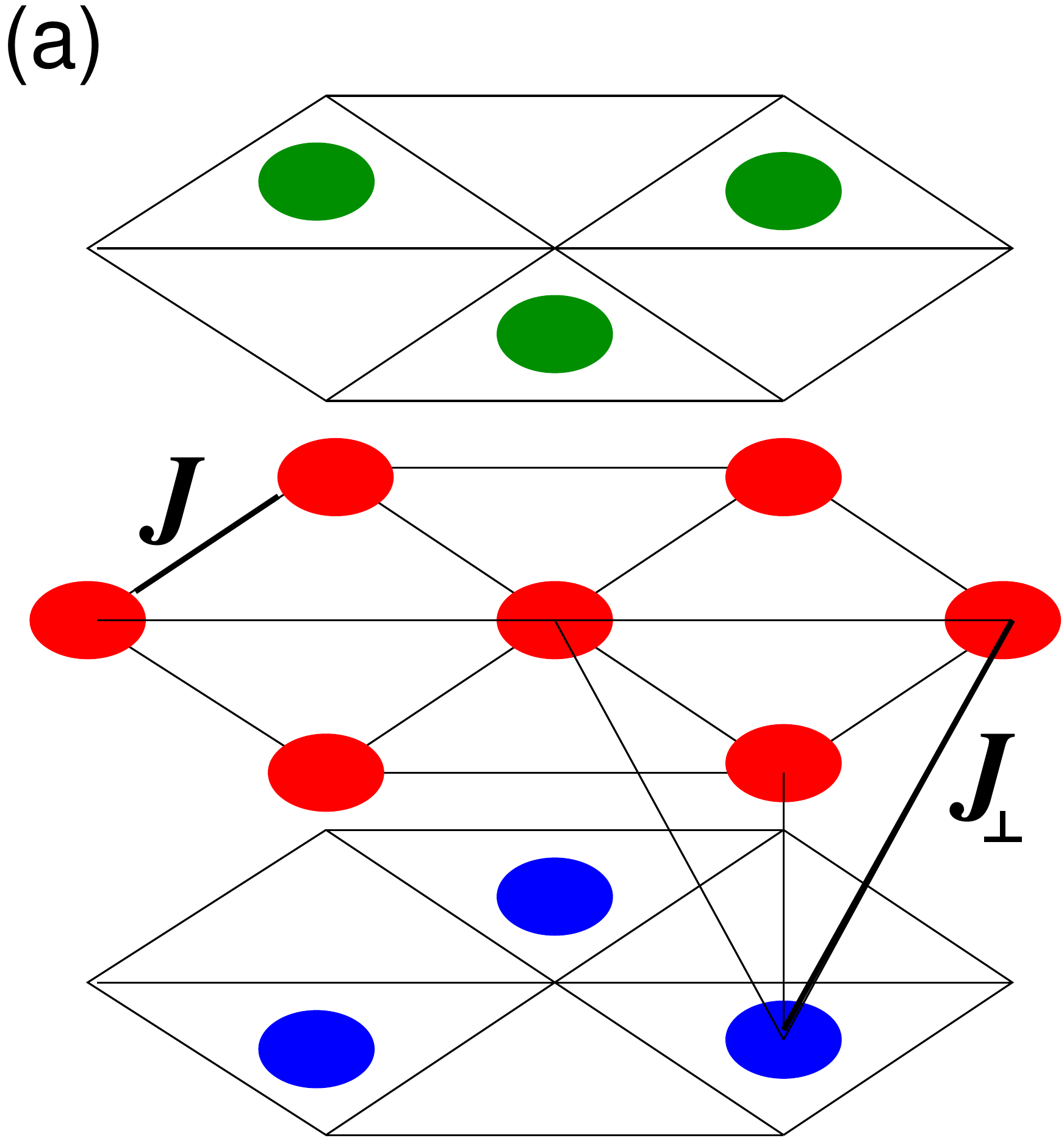} \hspace{0.2cm}\includegraphics[width=.4\linewidth]{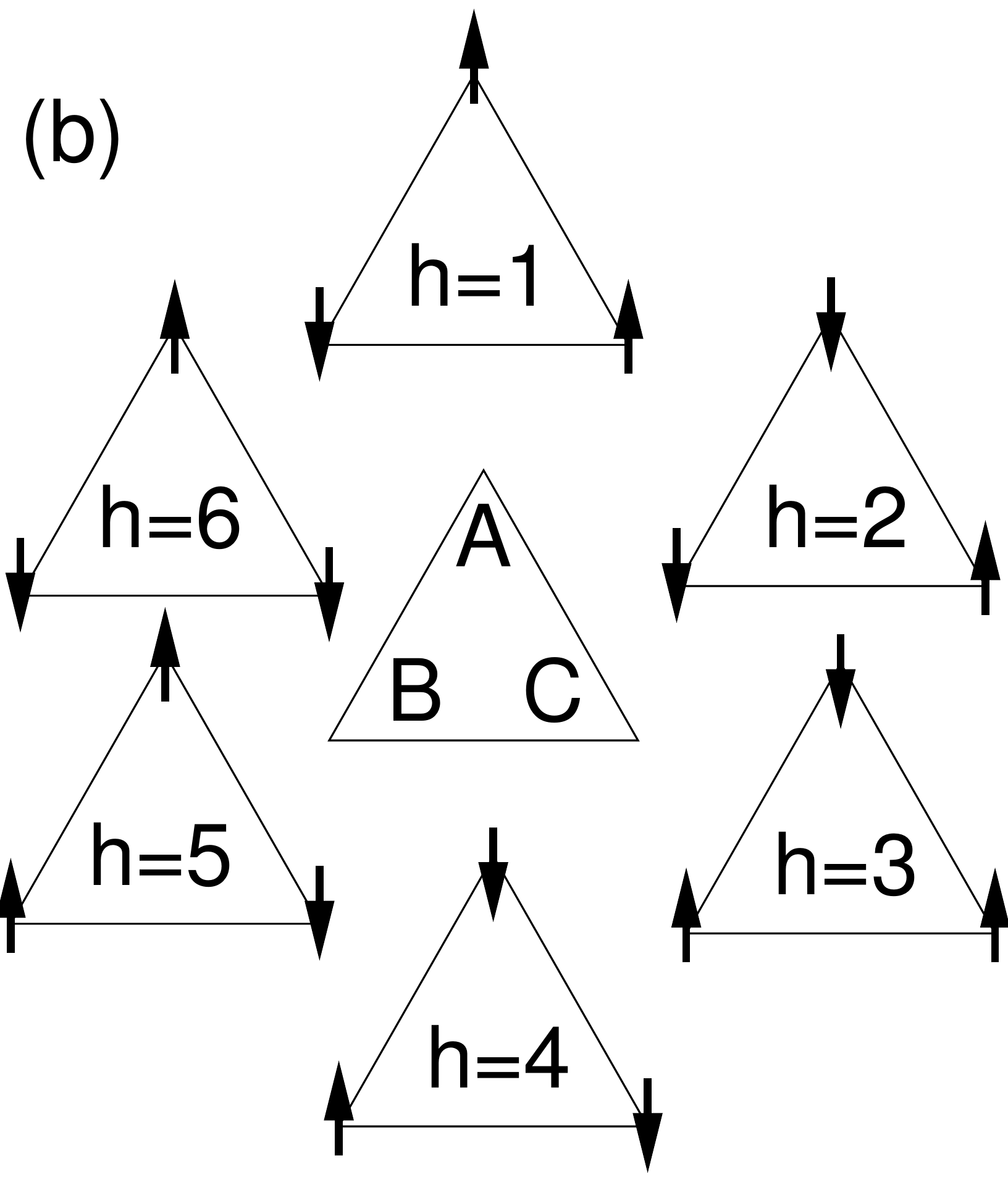} 
 \caption{(a) The multilayer triangular lattice with {\it abc} stacking.  (b) The mapping (modulo 6) from ground state spin configurations on sublattices A, B and C to integer-valued height variables $h$ (adapted from \cite{zeng}).}
\label{LatticeFig}
 \end{figure}

We start from the model illustrated in Fig.~\ref{LatticeFig}(a), with the Hamiltonian
\be \label{TheHamiltonian}
H = J  \sum_{\langle ij\rangle,z} \sigma_{i,z} \sigma_{j,z} + J_\perp \sum_{\{ij\},z}  \sigma_{i, z} \sigma_{j,z+1}\,. 
\ee
Here $\sigma_{i,z}=\pm1$, the integer $z$ labels planes, nn pairs of sites in the same plane are denoted by $ \langle ij \rangle$, and those from different planes by $\{ij\}$.  
Of the possible further-neighbour interactions that we have omitted, the most important is the unfrustrated coupling
\be
H_3 = J_3 \sum_{i,z} \sigma_{i, z} \sigma_{i, z+3} 
\ee
between a spin in layer $z$ and the spin directly above it in layer $z+3$, which lifts the degeneracy of the helical minima in ${\rm J}({\bf q})$.

To discuss the low-temperature behaviour of this model we exploit the fact that ground states of a single triangular layer can be described by a height model \cite{heightmodel,zeng}. 
Excitations out of the ground state 
are represented by screw dislocations or `vortices' in the height field. Vortex-antivortex pairs, present at any finite $T$, are unbound in an isolated layer and the vortex separation $\xi$ sets the correlation length, which is large at low $T$ because they are dilute.

Interlayer coupling greatly reduces the degeneracy. Ground states \cite{danielian} for $J_\perp \not=0$ are described by a height field with a gradient of maximum magnitude but arbitrary direction.
Of these, thermal fluctuations favour six states \cite{mackenzie} in which each triangular layer contains alternating stripes of up and down spins. For low $T$ the system 
adopts one of the three-dimensionally (3D) ordered, symmetry-breaking phases of this type.

At intermediate temperatures, thermal fluctuations compete with the interlayer coupling. For the regime of most interest, $J_\perp \ll T \ll J$, it is necessary to take two aspects of the physics into account. First, perturbing around a system of uncoupled layers, we use a renormalization-group (RG) calculation \cite{nienhuis} to treat the couplings $K_\perp\equiv J_\perp/T$ and 
$K_3 \equiv J_3/T$, which are RG-relevant.
Second, we make a detailed analysis of the influence of defects that appear in the coupled height model at finite temperature.
The outcome depends on three quantities: 
the vortex separation $\xi$; the scale $\ell_\perp$ at which the renormalised coupling $K_\perp\sim {\cal O}(1)$ in the vortex-free system; and the strength of the renormalised $K_3$. Interlayer coupling without vortices leads to 3D order at the scale $\ell_\perp$. In addition, it generates a potential that confines vortex-antivortex pairs. 
In many settings, unbound vortices destroy long-range order but bound vortex-antivortex pairs do not. Remarkably,  we find in this model that 3D order is disrupted even by bound pairs if 
$K_3$ is smaller than a critical value $K_3^*$. 
Hence three distinct regimes of behaviour emerge from this analysis, with: (a) weakly-coupled paramagnetic layers, when $\xi \ll \ell_\perp$; (b) 3D order, when $\ell_\perp \ll \xi$  and $K_3> K_3^*$; and (c) strong interlayer correlations but no long-range order, when $\ell_\perp \ll \xi$ and $K_3< K_3^*$.

We now set out these calculations in more detail. For a single triangular layer, the mapping \cite{heightmodel,zeng} between ground-state spin configurations and single-valued height variables, defined at the centres of triangles, is illustrated in Fig.~\ref{LatticeFig}(b). The inverse mapping has the form
$\sigma_\alpha=f(h+s_\alpha)$,
where $f(h+6)=f(h)$ with $f(h)=+1$ for $h=-1,0,1$ and $f(h)=-1$ for $h=2,3,4$. Here the sublattice label $\alpha=A, B$ or $C$ and $s_A=0$, $s_B=2$, $s_C=-2$. The dominant contribution to long-distance correlations involves the lowest Fourier component of $f(h)$, so that
\be \label{SCnst}
\sigma_\alpha \sim \cos \frac{\pi}{3}(h + s_\alpha).
\ee
Excitations out of the ground state consist of triangles in which all three spins have the same orientation, and the height field changes by $\pm 6$ around a closed path surrounding one such excitation.

The probability of a coarse-grained height configuration $h({\bf r})$ for an isolated layer is proportional to $e^{-{\cal H}_{\rm 2D}}$ with an effective Hamiltonian \cite{heightmodel,zeng}
\be \label{HeightHam}
{\cal H}_{\rm 2D} = \int d^2 {\bf r} \left \{ \frac{K}{2} \left|\nabla h ({\bf r})  \right|^2 - g \cos 2 \pi h({\bf r })  \right \}\,.
\ee
The first term in Eq.~(\ref{HeightHam}) describes the entropic cost of height gradients; the second term encodes the fact that microscopically the height field is integer-valued.  
The value of the stiffness $K$ is fixed by comparison with the known asymptotic behaviour $\langle \sigma_\alpha( {\bf r })  \sigma_\beta(0) \rangle \sim r^{-1/2} \cos{ \frac{\pi}{3} (s_\alpha- s_\beta)}$ of ground-state correlations in the Ising model \cite{Stephenson}: for $K = \frac{\pi}{9}$ this form is recovered and $g$ is RG-irrelevant. Note that, because of the sublattice-dependent phase $s_\alpha$ in Eq.~(\ref{SCnst}), small wavevector fluctuations of $h({\bf r})$ represent Ising spin fluctuations with wavectors near the corners of the triangular-lattice Brillouin zone.

To describe a multilayer system we introduce a height field $h_z({\bf r})$ in each layer, with ${\bf r} \equiv (x,y)$. Using (\ref{SCnst}), the inter-layer coupling written in terms of heights is \cite{supp}
\begin{widetext}
\be \label{InterLayerC}
\sum_{ \{ij\},z}  \sigma_{i,z} \sigma_{j,z+1} = -  \sum_z  \int \frac{d^2 {\bf r}}{\ell} \left( \partial_x  h_z({\bf r})    \cos \frac{ \pi}{3} (h_{z+1}({\bf r}) - h_z({\bf r}) )
- \partial_y  h_z({\bf r}) \sin \frac{ \pi}{3} (h_{z+1}({\bf r}) - h_z({\bf r}) )   + ...\right)
\ee
where $\ell$ is the short-distance cut-off and the ellipsis indicates omitted terms that are RG-irrelevant at weak interlayer coupling. We introduce the notation $\delta^{(p)}_z ({\bf r}) = \frac{\pi}{3} \left( h_{z+p} ({\bf r}) - h_z ({\bf r})  \right)$ and define a reduced interlayer coupling $\kappa_\perp={K_\perp}/{\ell K}$. The leading contributions to the multilayer height Hamiltonian can then be combined as
\be \label{HeightHam2}
{\cal H}_{\rm 3D} =\frac{K}{2} \sum_z\int d^2 {\bf r}  \left \{ \left(\partial_x h_z ({\bf r})  - \kappa_\perp \cos \delta^{(1)}_z ({\bf r} )  \right)^2 + \left(\partial_y h_z ({\bf r} )  + \kappa_\perp \sin \delta^{(1)}_z ({\bf r} )  \right)^2 -\kappa_\perp^2 \right\}.
\ee
\end{widetext}

This model has a striking continuous symmetry under a joint real-space and height-space transformation. Let $R_\theta$ denote rotation in the $xy$ plane by the angle $\theta$. Then the transformation ${\bf r}^\prime = R_\theta({\bf r})$ and
\be
h_z({\bf r}) \to h^\prime_z({\bf r}) = h_z({\bf r}^\prime) + 3\theta z/\pi
\ee 
leaves ${\cal H}_{\rm 3D}$ invariant for any $\theta$. 
Ground states of ${\cal H}_{\rm 3D}$ 
are parameterised by constants $\theta$ and $\alpha$, having a form
\be \label{GStatesEq}
h_z^{(\theta)}({\bf r}) = \kappa_\perp (x \cos \theta - y \sin\theta) + 3\theta z/\pi +\alpha
\ee
that balances interlayer coupling energy against intralayer entropy.

Additional interactions are expected to restrict the continuous symmetry to the discrete one ($\theta = 2\pi n/3$ with integer $n$) of the microscopic model. Of these locking terms, the most important are the RG-relevant ones that link nearby layers, and the dominant
one, $K_3 \cos \delta^{(3)}_z ({\bf r})$, acts between layers with separation three.
All locking terms that act between nearest-neighbour layers are RG-irrelevant; the leading example is
$K_1[(\partial_x  h_z({\bf r}))^2 - (\partial_y  h_z({\bf r}))^2]  \cos \delta^{(1)}_z ({\bf r}) + 2(\partial_x  h_z({\bf r}))(\partial_y  h_z({\bf r})) \sin \delta^{(1)}_z ({\bf r})$.

We compute the RG flow \cite{nienhuis} of the interlayer couplings $K_\perp$ and $K_3$ and the vortex fugacity $y$ as a function of 
$\ell$ near the Gaussian fixed point of the uncoupled multilayer system. 
To leading order in the couplings
\begin{eqnarray} \label{RGFlows}
\frac{\partial \ln K_\perp}{ \partial \ln \ell } & =&
\left ( 1- \frac{\pi}{18 K} \right ) \n
\frac{\partial K_3}{ \partial \ln \ell } & =& K_3 \left ( 2- \frac{\pi }{18 K} \right ) \n
{\rm and} \quad \frac{\partial y}{ \partial \ln \ell } & =& y \left ( 2-  \frac{9  K}{  \pi} \right )\,.
\end{eqnarray}
Higher order contributions generate $K_3$ (initially zero in a model with only nn interactions) from $K_\perp$ and $K_1$, and also renormalise $K$, but are otherwise unimportant at small coupling.

This description of the RG flow breaks down at the scale where the largest of $K_\perp$, $K_3$ and $y$ is ${\cal O}(1)$.
If  $y\sim 1$ with $K_\perp, K_3$ $\ll1$, the RG treatment of weakly coupled layers can be applied at all scales: for $y \gtrsim 1$, $K$ flows to zero, and hence so do $K_\perp$ and $K_3$, yielding a conventional paramagnet. Alternatively, when $K_\perp \sim 1$ with $y \ll1$, layers are strongly coupled at large scales and the state of the system depends on a competition between vortex and locking effects, which we now examine.

To discuss the system of strongly coupled layers, we consider the energy cost of small amplitude fluctuations about a ground state of ${\cal H}_{\rm 3D}$. Let 
\be\label{decomposition}
h_z({\bf r})= h_z^{(\theta)}({\bf r}) + \varphi_z ({\bf r})
\ee
and write (with ${\bf q}_\perp =(q_x,q_y)$)
\be
\varphi_z ({\bf r}) = \frac{1}{(2\pi)^3}\int {\rm d}^3 {\bf q}\, \varphi({\bf q}) e^{i({\bf q}_\perp {\bf r} + q_z z)}\,.
\ee
Then at quadratic order
\be
{\cal H}_{\rm 3D} =  \frac{K}{2(2\pi)^3}\int {\rm d}^3 {\bf q}\, {\cal E}({\bf q})|\varphi({\bf q})|^2
\ee
with (taking $\theta=0$)
\be\label{EpsEq}
{\cal E}({\bf q})  = q_x^2 + (q_y + \kappa_\perp \sin q_z)^2 + \kappa_\perp^2(1-\cos q_z)^2 \, .
\ee
This dispersion relation is unusually soft ($\propto q_z^4$) in the interlayer direction. (The fluctuation energy for smectic liquid crystals \cite{CL} has a similar form, but with two quartic directions and a single quadratic one.) Because $\langle \varphi^2_z ({\bf r}) \rangle$ computed Eq.~(\ref{EpsEq}) is finite, the scaling flow of Eq.~(\ref{RGFlows}) stops at $K_\perp \sim 1$. Without vortices, the system has long-range order and any non-zero locking interaction pins $\theta$.

To understand the influence of vortices on the system at $K_\perp \gtrsim 1$ we should examine ground states of ${\cal H}_{\rm 3D}$ in height-field sectors with fixed vortex locations. The outcomes are, first, confinement of vortex-antivortex pairs, and second, destruction of long-range order by bound pairs if $K_3$ is small. 

Consider introducing a single vortex-antivortex pair in layer $z=0$. Then, using the notation of Eq.~(\ref{decomposition}), $\varphi_0({\bf r})$ is multiple-valued, winding by $6$ $(-6)$ along any closed path encircling the vortex (anti-vortex). 
At large vortex separation 
$\varphi_0({\bf r})$ has a step of height 6 and width $w$ on the line joining the vortex centres. The energy cost per unit length of this step is 
\be
\varepsilon \sim Kw(w^{-2} + \kappa_\perp^2) \propto K\kappa_\perp\,
\ee
where the last expression follows from minimising over $w$ and the optimal width is $w\sim\kappa_\perp^{-1}$. Hence vortices at separations large compared to $\kappa_\perp^{-1}$ are subject to a linear confining potential, and pairs are tightly bound.

Tightly-bound pairs generate distortions in $h_z({\bf r})$ that fall off only slowly with distance from the pair centre. The appropriate far field can be induced without considering a multiply-valued $\varphi_0({\bf r})$ by instead adding a coupling $-K\int {\rm d}^2{\bf r}\, v({\bf r})\varphi_0({\bf r})$ to ${\cal H}_{\rm 3D}$. We find \cite{supp} that the Fourier transform of the required potential $v({\bf r})$ has the form
\be
v({\bf q}) = 6i  \,\hat{\bf z}\cdot({\bf q}\times {\bf b})
\ee
in the limit $q\ll \kappa_\perp$, for a pair that has separation vector $\bf b$ and its centre at the origin. In the minimum-energy state containing this pair, the single-valued far field at ${\bf r},z$ is (with ${\bf q}_\perp = (q_x,q_y)$)
\be
\varphi_z({\bf r})=\frac{1}{(2\pi)^3} \int {\rm d}^3 {\bf q}\,  \frac{ v({\bf q})}{{\cal E}({\bf q})} 
e^{i({\bf q}_\perp{\bf r} + q_zz)} \,.
\ee

Extending this calculation to randomly located pairs at density $\rho$, we find that they generate fluctuations in $\varphi_z({\bf r})$ with mean square amplitude
\be\label{squareflucs}
\langle [\varphi_z({\bf r})]^2\rangle = \frac{\rho}{(2\pi)^3} \int {\rm d}^3 {\bf q}\,  \frac{ |v({\bf q})|^2}{{\cal E}^2({\bf q})}\,.
\ee
The integral on the right of Eq.~(\ref{squareflucs}) is divergent at small ${\bf q}$, indicating that any non-zero density of bound vortex pairs destroys long-range order in the absence of locking interactions. The in-plane and interlayer correlation lengths $\xi_\perp$ and $\xi_z$ can be estimated by using finite system size as a cut-off. Taking $\langle |{\bf b}|^2\rangle \sim \ell^2$, we obtain the highly anisotropic results $\xi_z \sim (\rho \ell^2 )^{-1}$ and $\xi_\perp \sim \kappa_\perp^{-1}(\rho \ell^2)^{-2}$. 

Spin correlations at wavevectors that have in-plane components close to one of the corners ${\bf K}$ of the triangular lattice Brillouin zone can be expressed in terms of small-${\bf q}_\perp$ components of $h_z({\bf r})$. From Eq.~(\ref{SCnst}) we find \cite{supp}
\begin{eqnarray}
 S({\bf q})&\equiv&\sum_{jz}  \,e^{i({\bf K}+{\bf q}){\bf r}_{jz}} \langle \sigma_{00}\sigma_{jz}\rangle\\
 &\sim&\sum_z \int {\rm d}^2{\bf r} \, \,e^{i({\bf q}_\perp{\bf r}+[q_z+\frac{2\pi p}{3}] z)} \left\langle e^{\pm i\frac{\pi}{3}[h_z({\bf r})-h_0({\bf 0})]}\right\rangle\nonumber
\end{eqnarray}
where, on the right-hand side, the choice of $\pm$ sign and the value of the integer $p$ depend on the zone corner.

Correlations on scales shorter than $\xi_z$, $\xi_\perp$ resemble those in ground states and can be calculated from 
$h_z^{(\theta)}({\bf r})$ [Eq.~(\ref{GStatesEq})] by averaging over $\theta$. The result of this approximation is a sharply defined helix
\be
S({\bf q}) \propto \delta(q_x-\kappa_\perp\frac{\pi}{3}\cos q_z)\,\delta(q_y + \frac{2\pi p} {3}\pm \kappa_\perp\frac{\pi}{3}\sin q_z)\,,
\ee
which is broadened when finite $\xi_z$, $\xi_\perp$ are taken into account.

Locking interactions suppress fluctuations in $\varphi_z({\bf r})$ and stabilise long range order if their effect integrated over the correlation volume is large at the RG scale on which $K_\perp\sim1$. The condition $K_3 (\xi_\perp/\ell)^2\xi_z \sim 1$ implies a critical value of $K_3^{*}\sim (\rho \ell^2)^5$.
Varying temperature in the spin system, a regime with strong interlayer correlations separates the high-temperature paramagnet from the 3D ordered phase if the renormalised $K_3 \ll K_3^*$ at the scale for which $K_\perp\sim1$. 

Since, from Eq.~(\ref{RGFlows}), $K_3$ grows faster under RG than $K_\perp$, the existence of an intermediate regime requires both a sufficiently small microscopic value of $J_3$ and sufficiently slow generation of $K_3$ from $K_\perp$ and the RG-irrelevant coupling $K_1$. 
Because $K_3$ couples sites three layers apart, its value after an ${\cal O}(1)$ RG rescaling in a system that, at the microscopic level, has only nearest-neighbour interactions cannot be larger than ${\cal O}(J_\perp/T)^n$ with $n=3$. We find however (as noted in Ref. \onlinecite{Starykh} for a similarly frustrated 2D problem) that this leading-order contribution is absent and the lowest non-vanishing contribution has $n=7$. 

To obtain a phase diagram for the system, we integrate the RG flow equations (\ref{RGFlows}) from a microscopic scale $\ell_0$ to the scale $\ell_\perp$ at which $K_\perp(\ell_\perp)=1$, with initial values $K_\perp=J_\perp/T$, $K_3\sim (J_\perp/T)^7$ and $y(\ell_0)=e^{-4J/T}$. The crossover between the conventional, high-temperature paramagnet and the cooperative paramagnet is at $y(\ell_\perp)\sim1$, implying $J_\perp \sim T\big[y(\ell_0)\big]^{1/2}$. The ordering transition is at $K_3(\ell_\perp)=K_3^*$; setting $\rho\ell^2 = \big[y(\ell)\big]^2$ in the expression derived above for $K_3^*$, this implies $J_\perp \sim T\big[y(\ell_0)\big]^{5/12}$. The cooperative paramagnet therefore occupies the range of interlayer couplings
\be
T\big[y(\ell_0)\big]^{1/2} \lesssim J_\perp \lesssim T\big[y(\ell_0)\big]^{5/12}\,.
\ee
In the small $J_\perp/J$ limit of interest, this simplifies: the onset of strong interlayer correlations is at  $J_\perp \approx Je^{-2J/T}$, while the ordering transition is at the much smaller value $J_\perp \approx  Je^{-5J/3T}$. Correlations lengths in the cooperative paramagnet are given by $\xi_z \approx e^{8J/T} (J_\perp/J)^4$ and $\xi_\perp \approx \ell_0 (J/J_\perp)^2 \xi_z^2$. Hence the interlayer correlation length increases from $\xi_z\approx 1$ at onset to $\xi_z \approx (J/J_\perp)^{4/5}$ near the ordering transition, and the in-layer correlation length from $\xi_\perp \approx \ell_0(J/J_\perp)^2$ to $\xi_\perp \approx \ell_0(J/J_\perp)^{18/5}$. The parametrically large values reached by the correlation lengths are confirmation that the system indeed behaves as a cooperative paramagnet.

In summary, we have discussed the nearest-neighbour Ising antiferromagnet  on a lattice of weakly-coupled triangular layers with a frustrated {\it abc} stacking, showing that it has a low-temperature regime in which there are strong interlayer correlations without long-range order. This cooperative paramagnet is striking both for its correlations, which generate maxima in $S({\bf q})$ on helices in reciprocal space, and for the mechanism of order-suppression via vortex-antivortex bound pairs, which is unlike that in any other system we are aware of.
We expect that weakly coupled triangular layers in an hcp lattice (the {\it abab} stacking) \cite{hcp1,hcp2} will exhibit a similar cooperative paramagnetic regime.
This system, together with a numerical study of the model with {\it abc} stacking, will be discussed in detail in forthcoming work \cite{forthcoming}.

An Ising model similar to Eq.~\ref{TheHamiltonian} has been proposed \cite{yamada,harris} as a description of charge ordering in the materials LuFe$_2$O$_4$ and YbFe$_2$O$_4$: Ising pseudospins represent the charge states Fe$^{2+}$ and Fe$^{3+}$, and antiferromagnetic coupling arises from Coulomb repulsion.  Experimental studies \cite{yamada,radaelli,review}, in particular of  YbFe$_2$O$_4$  \cite{radaelli}, find helices of scattering intensity in a temperature range above a three-dimensional charge-ordering transition. While an accurate description of these materials would require treating both a more complicated (bi-layer) structure and additional (spin) degrees of freedom, the results we present in this paper elucidate how strong interlayer correlations can arise without long-range order.


We thank F. H. L Essler, O. A. Starykh and especially P. G. Radaelli for discussions. FJB is supported by NSF-DMR 1352271. JTC is supported in part by EPSRC Grant No. EP/I032487/1.

\vfill
\newpage

\centerline{\bf Supplementary material}

\vspace{0.2cm}
We provide derivations of the inter-layer coupling in terms of height fields, the far-field potential introduced by bound vortex pairs, and the  structure factor in the cooperative paramagnet regime.  

\section{Height mapping and inter-layer coupling}

As first observed in Ref. \onlinecite{heightmodel}, the possible ground state configurations in a triangular lattice Ising antiferromagnet can be represented by a height field.  The height representation we use here follows Ref. \onlinecite{zeng}, in which a height is assigned to the centre of each triangle.  To do so, we choose a convention for dividing the triangular lattice into three sublattices $A$, $B$, and $C$.  The six possible ground-state spin configurations on a given triangle can then be identified with 6 physically distinct height variables as follows:

\begin{align} \label{Eq_Hspins}
h \ \ \  \ \ \ & (\sigma_A ,\ \sigma_B, \ \sigma_C )& h \ \ \  \ \ \ & (\sigma_A ,\ \sigma_B,\ \sigma_C) \n
0\ \ \  \ \ \ &  (+,\ - , \ - ) & 1\ \ \   \ \ \ & ( +,\  -, \ + ) \n
2 \ \ \  \ \ \ &(  - ,\ - ,\ + ) &
3\ \ \   \ \ \ & ( - ,\ + ,\ + )\n
4 \ \ \  \ \ \ & ( - ,\ + ,\ - ) &
5 \ \ \  \ \ \ &( + ,\ + ,\ -) 
\end{align} 
The height field is single-valued provided that each triangle contains no more than one frustrated bond.  
Spin configurations containing triangles in which spins are all up or all down correspond to vortices in the height field.

To obtain the inter-layer couplings, we work with the inverse mapping, given by
$\sigma_{\alpha}   = f (h + s_\alpha)$, with $(s_A, s_B, s_C) =(0,2,-2)$, and 
\begin{align} \label{Spin2H}
f(h)   = \frac{4 }{3} \cos \frac{ \pi h}{3} - \frac{1}{3} \cos \pi h
\end{align}
Provided $h$ is an integer, this reproduces exactly the spin configurations in Eq. (\ref{Eq_Hspins}), irrespective of which of the 6 triangles neighbouring the site in question is used.  
In order to coarse-grain the system, we will use the convention
$\sigma_{z, \alpha} ({\bf r})  = f(h_z ({\bf r} + {\bf a}_\alpha )+  s_\alpha )$, where 
\be
{\bf a}_A = \left( 0, \ -\frac{1}{\sqrt{3}} \right )  \ \ \ {\bf a}_B = \left( \frac{1}{2}, \ \frac{1}{2\sqrt{3}} \right )   \ \ \ {\bf a}_C = \left( -\frac{1}{2}, \ \frac{1}{2\sqrt{3}} \right )  
\ee

On the {\it abc}-stacked triangular lattice, the inter-layer Ising interaction couples each spin to three spins in each adjacent layer (one from each sublattice).
In terms of the height fields, this gives:
\begin{align}
 {\cal H}_{\perp}= J_\perp  \sum_{\alpha = A,B,C} \sum_{{\bf r}} f (h_z({\bf r}+ {\bf a}_\alpha )+ s_\alpha) \overline{f}({\bf r}, z+1)
\end{align}
where 
\be
\overline{f}({\bf r}, z) = \sum_{\alpha= A,B,C} f (h_{z+1} ({\bf r} + {\bf a}_\alpha )+ s_\alpha)
\ee
Since we are concerned with the interlayer coupling only for the long-wavelength components of the height fields, we may Taylor expand this expression in derivatives of $h_z({\bf r})$.  
The leading-order term is $ \cos \pi h_z \cos \pi h_{z+1}$
which is highly irrelevant for weak inter-layer couplings.
The only RG-relevant contribution, which comes from the first derivatives of $h$, is 
\begin{align}
{\cal H}_\perp& = & 
- \frac{ 4 \pi J_{\perp} }{9 \sqrt{3} } \sum_{{\bf r}, z} \left( \cos \frac{ \pi}{3} (h_{z+1} ({\bf r})- h_z({\bf r}) ) \partial_x  h_z ({\bf r}) \right . \n
&& \left. -  \sin \frac{ \pi}{3} (h_{z+1} ({\bf r})- h_z({\bf r}) ) \partial_y  h_z({\bf r}) \right ).    \n
\end{align}
Approximating the sum by an integral (and omitting the factor $4\pi/9\sqrt{3}$ for simplicity) reproduces the expression given in the main text.

\section{ Effective potential due to bound vortex pairs}

A key aspect of our argument for the existence of a cooperative paramagnetic regime is the fact that, due to the unusually soft dispersion along the stacking direction, even tightly bound vortex pairs are sufficient to frustrate long-ranged order in our model.  Here we derive the form of the effective potential $v({\bf q})$ used to demonstrate this fact in the main text.

We consider the far-field potential from a vortex-antivortex pair in an isolated layer.  The vortex and antivortex are separated by a vector ${\bf b}$. We take the pair to be centered at the origin.  This pair is described by the height field configuration
\be
h
(x,y) = \frac{3}{ \pi} \left[ \arctan \left( \frac{ x + {\bf b} /2 \cdot \hat{x}}{y+{\bf b} /2 \cdot \hat{y}} \right) -  \arctan \left( \frac{ x - {\bf b}/2  \cdot \hat{x}}{y- {\bf b}  /2\cdot \hat{y}} \right)   \right]
\ee
For $|{\bf r}| \gg |{\bf b}|$, we have:
\be
h
(x,y) \approx \frac{3}{ \pi} \frac{\hat{z} \cdot ( {\bf b} \times {\bf r} )}{r^2 }
\ee
or, equivalently,
\be \label{FarFieldh}
h({\bf q}) \approx 6 i \frac{ \hat{z} \cdot ({\bf q} \times{\bf b}  ) }{q^2 } 
\ee

The far-field height configuration in Eq.~(\ref{FarFieldh}) can be induced by adding a potential term $v{(\bf q})$ to the effective Hamiltonian for the height field.  Specifically, for an isolated layer, the effective Hamiltonian  $ (K/[2\pi]^2) \int d^2 {\bf q} \left [  \frac{1}{2} {\mathcal E} ({\bf q})  | \varphi_z({\bf q}) |^2  - \varphi_z(-{\bf q}) v_z({\bf q})  \right]$ has the minimum energy configuration
\be
\overline{\varphi}({\bf q})_z  = \frac{ v_z({\bf q})}{{\mathcal E}({\bf q}) } =  \frac{ v_z({\bf q})}{{q}^2 }
\ee
Thus choosing a potential
\be
v({\bf q}) = 6 i \hat{z} \cdot ({\bf q} \times  {\bf b} ) 
\ee
we recover the correct far-field configuration.  As shown in the main text, this potential term destroys the long-ranged order in $\varphi$.

\section{Correlation functions from height variables}

Here we derive the expression for the spin-spin correlation function in terms of the height fields.   
Let the primitive lattice vectors within a triangular layer be ${\bf a}_1' = (1,0)$ and ${\bf a}_2'=(1/2,\sqrt{3}/2)$. 
Then the primitive lattice vectors of the {\it abc}-stacked triangular lattice have the form
\be
{\bf a}_1 = ({\bf a}_1' ,0)\ , \ \ {\bf a}_2 = ({\bf a}_2' ,0)\ , \ \ \ {\bf a}_3 = \left( \frac{2}{3} {\bf a}'_2 -\frac{1}{3} {\bf a}'_1,1  \right )
\ee
where we have taken the inter-layer distance to be $1$ for convenience. Let ${\bf b}_1$, ${\bf b}_2$ and ${\bf b}_3$ denote the corresponding reciprocal lattice vectors.

We first observe that the corners of the triangular lattice Brillouin zone lie at
\be
{\bf K} = \frac{2}{3} {\bf b}_1 + \frac{1}{3} {\bf b}_2 \ , \ \ \ {\bf K}' = \frac{1}{3} {\bf b}_1 + \frac{2}{3} {\bf b}_2 
\ee 
Therefore, for a given position ${\bf r} = n_1 {\bf a}_1 + n_2 {\bf a}_2+ z {\bf a}_3$ on the stacked triangular lattice, we have
\begin{align}
\left( {\bf K} + h_1 {\bf b}_1 + h_2 {\bf b}_2 \right ) \cdot {\bf r} = \frac{4 \pi }{3} ( n_1 + h_2 z) + \frac{2 \pi }{3} ( n_2 - h_1 z ) \n
\left( {\bf K}' + h_1 {\bf b}_1 + h_2 {\bf b}_2 \right ) \cdot {\bf r} = \frac{2 \pi }{3} ( n_1 - h_1 z) + \frac{4 \pi }{3} ( n_2 + h_2 z )
\end{align}
modulo $2 \pi$.  
The combination $\frac{2 \pi }{3} ( 2 n_1+ n_2)$  mod $2 \pi$ is simply $\frac{ \pi }{3}s_\alpha$. Therefore
\begin{align}
e^{i \left( {\bf K} + h_1 {\bf b}_1 + h_2 {\bf b}_2 \right ) \cdot {\bf r} } = e^{i  \pi ( s_\alpha  + 2pz )/3 } \n
e^{i \left( {\bf K}' + h_1 {\bf b}_1 + h_2 {\bf b}_2 \right ) \cdot {\bf r} } = e^{i  \pi ( - s_\alpha  +2p z  )/3}
\end{align}
where the integer value of $p=(2 h_2 - h_1)$ depends on the Brillouin zone corner under consideration. Now, the most relevant terms in the spin-to-height mapping are
\be
\sigma_\alpha ({\bf r}) \sim  \left( e^{ i  \pi/3 \ h({\bf r})} e^{ i   \pi/3 s_\alpha} + e^{ - i  \pi/3 \ h({\bf r})} e^{ - i   \pi/3 s_\alpha}\right)\,.
\ee
Keeping only slowly varying terms, we therefore obtain
\begin{widetext}
\begin{align}
\sum_{i,z} \,\sigma_{i,z} \sigma_{0,0} e^{ i ( {\bf K} +{\bf q} )\, \cdot{\bf r} + z q_z} \approx   \sum_z \int d^2 {\bf r}\, e^{ - i  \frac{\pi}{3} \left( h ({\bf r} )  - h ({\bf 0} ) \right )  }  e^{i z(\frac{2 \pi}{3}  p +  q_z) } e^{ i {\bf q} \cdot{\bf r} } 
\end{align}
\end{widetext}
which gives Eq.~(18) of the main text.

\end{document}